\RequirePackage{fix-cm}
\documentclass[smallextended]{svjour3}       
\smartqed  
\usepackage{graphicx}
\usepackage{latexsym}
\usepackage{amsmath,amssymb}
\usepackage{caption}

\def\bd{\begin{displaymath}}
\def\be{\begin{equation}}
\def\ed{\end{displaymath}}
\usepackage[normalem]{ulem}
\usepackage{soul}
\def\ee{\end{equation}}

\begin{document}

\title{Decoherence due to thermal effects in two quintessential quantum systems}



\author{S. Nussinov,  T. Madziwa-Nussinov,  \\
and  Z. Nussinov}


\institute{S. Nussinov \at
              Raymond and Beverly Sackler School of Physics and Astronomy\\
              Tel Aviv University, Tel Aviv, Israel\\
              and    \newline
              Institute for Quantum Studies, Chapman University\\
              Orange, CA 92866\\
           \and
           T. Madziwa-Nussinov  and Z. Nussinov \at
              Physics Department\\
              Washington University \\
              St. Louis, MO, 63130, USA
}

\date{Received: date / Accepted: date}

\maketitle

\begin{abstract}
Decoherence effects at finite temperature ($T$) are examined
for two manifestly quantum systems: (i) Casimir forces between parallel plates that
conduct along different directions, and (ii) a topological Aharonov-Bohm (AB) type 
force between fluxons in a superconductor.  As we illustrate, standard path integral calculations
suggest that thermal effects may remove the angular dependence of the Casimir force in case (i)
with a decoherence time set by $h/(k_{B} T)$ where $h$ is Planck's constant and $k_{B}$ is
the Boltzmann constant. This prediction
may be tested. The effect in case (ii) is due a
$\pi$ phase shift picked by unpaired electrons upon encircling an
odd number of fluxons. In principle, this effect may lead to small modifications in Abrikosov lattices. While the AB forces exist at extremely low temperatures,
we find that thermal decoherence may strongly suppress the
topological force at experimentally pertinent finite temperatures.  It is suggested that both cases (i) and (ii) (as well as other
examples briefly sketched) are related to a quantum version of the
fluctuation-dissipation theorem.
\end{abstract}

\section{Introduction}
\label{intro}

\noindent A quantum system in equilibrium at a finite temperature is not in
a pure state, but rather is described by a thermal density matrix. However,
 just like a classical pendulum or a loaded spring which keep performing
for some time their periodic motion in a thermal environment, so can an
appropriately entangled quantum system keep its coherence for a while. The
thermal environment generates a random fluctuating force operating on the
classical harmonic oscillator system. The 
``fluctuation-dissipation theorem'' (see, e.g., \cite{Kubo1990})  then implies that
these damp the initial ordered periodic motion leaving only the $O(k_{B}T)$
thermal energy per degree of freedom. Similarly, thermal effects will tend
to decohere the quantum system, which started say as a pure state. \cite{huse}

Rather than address the full and involved question of the general quantum
version of the fluctuation dissipation theorem, we address here two
examples. These are two effects which to certain degrees are spoiled and
weakened by a randomizing and decohering thermal milieu. While these
effects seem superficially to be unrelated, we believe that at a deeper
level they are and both pertain to the general issue considered.

The first effect is a ``polarized" version of the Casimir force between two
neutral parallel conducting plates \cite{Nussinov0000}. It arises when the
two plates conduct in different directions. Beyond its $1/{a^4}$ dependence
on the separation $a$ of the two plates, the Casimir force per unit of area
then depends also on the angle $\beta$ between the directions of
conductance \cite{Kenneth1998}. This dependence reflects the vector nature
of light, namely, the polarization degree of freedom. The Casimir force
and its variant can be viewed, like many other forces, as being generated
by the the interaction of the plates (or ``big" objects in general) with
smaller elements or modes of the intervening medium. The forces can be
calculated exactly by using the euclidean path integral method introduced
and extensively used in this context in the thesis of O. Kenneth \cite{Kenneth_thesis}.

The Casimir force per unit area at finite temperatures is well known
\cite{Milton2001} and  can be readily derived by the path integral approach
where the time direction is compactified down to a size of $1/(k_{B}T)$.
 When $k_{B}T>1/a $, the thermal Casimir force has a $k_{B}T/{a^3}$
dependence and is stronger than the ordinary (i.e., zero temperature) 
Casimir force $-$ as indeed
suggested by a simplistic argument.

Below, we evaluate the polarized variant of the Casimir effect for finite
temperature. This calculation was motivated by the prospect that the effect
will be measured experimentally \cite{Markle2013}. The setup considered was
at room temperature of  300 K, i.e., $k_{B}T \sim eV/{40}$ and plate
separation a of few microns. Contrary to ``naive expectations",  the
calculation implies that the dependence of the force on $\beta$, the angle
between the conduction directions of the plate falls, off exponentially
with temperature like $\exp(-2 \pi k_{B} Ta/(\hbar c))$. This is attributed to the
randomizing effect on the polarization of the photons in a thermal
background as they travel between the two plates. 

The second effect concerns the ``Aharonov-Bohm Force" suggested to occur 
\cite{Aharonov0000a} between magnetic fluxons immersed in a fluid
of electrons due to phases which the electrons pick when circulating the
fluxons. In particular,  for ``semions",  namely half-fluxons - in units
of $hc/e$ where $e$ is the \textit{electron} rather than the Cooper pair
charge of 2$e$- which naturally arise inside superconductors, this
`topological' force was found to be attractive even for fluxons with
parallel magnetic fields. This was inspired on the one hand by analogy with
the Casimir effect with the electrons (rather than the vacuum and/or
thermal photons) playing the role of the relevant excitations of the medium
and by some remarkable features of the the ground state of electronic
system in the presence of half-fluxons \cite{Aharonov0000b}. To achieve it
in a real setup of physical superconductors, we have to use a finite
temperature $T \le{T_c}$ so that we have both supeconductivity and fluxons and a
finite fraction of unpaired electrons to generate the force of the form $F
\sim  n \hbar^2/m_e r $ with $r$ the separation between the fluxons and $n$
the two-dimensional number density of the electrons.  In Ref. \cite{Aharonov0000a}, 
it was remarked that unless somehow screened by countering currents,
the force was rather large.

Here we note that decohering effects can randomize the elctrons' phases
over time intervals of order $1/{k_{B}T}$ and naturally lead to an exponentially
falling force between two nearby semions. The reduced symmetry of the
present problem relative to the case of the Casimir parallel infinite
plates, hinders exact evaluation of the thermal path integral though it
suggests the same qualitative results.  We next speculate on the
feasibility of testing the effect now that we have it in a more realistic
and correct form.

The plan of the paper is as follows: Section 2 presents a discussion of the
rather long relaxation time of a simple classical harmonic motion in
certain cases even in the presence of a thermal background. In sections 3
and 4 we briefly review the two effects/forces mentioned above and present
the new and more involved finite temperature discussion. In section 5 we
comment on the connection with the general theme of thermal decoherence and
conclude.


\section{A Short Note on Classical
Decoherence}
\label{sec:2}  

It is instructive to recall how classical systems can maintain ``coherent"
motion despite a thermal background $-$ as we conjecture that similar
persistence effects may be relevant also to the quantum case. Consider then
an `ideal' sensitive torsion balance designed to measure tiny $-$ say,
gravitational forces. It consists of a horizontal bar hung at its center by
a thin long and ideally(!!) a loss-free fiber. The system is enclosed
inside a cylinder pumped to very low pressure of  $\sim10^{-10}$
atmospheres. When gravitationally torqued by a pair of massive near-by
spheres, the bar and fiber system starts performing a periodic harmonic
twist motion.

The question of interest is the extent to which finite, say even room
temperature, effects of the medium impair the function of the torsion
balance. For generic $O(0.1)$ kg masses of rod and external spheres,
dimensions and distances of $O(10)$ cm, and a nano-radian rotation of the
bar, the gravitational energy involved and that of the resulting energy in
the oscillatory mode are comparable with or smaller than $k_{B}T$, the
generic thermal energy per degree of freedom. However, only after that much
energy is pumped  to that particular mode of motion of the rod will it's
motion be appreciably affected. After the system relaxes to thermal
equilibrium - all degrees of freedom and that of the whole rod considered
here as well, will by the equipartition theorem, each have an energy of $k_{B} T$. However, if this relaxation is driven only by collisions
with the ambient residual gas molecules (or with thermal photons which only
drastic cooling can eliminate) the times required for damping this motion
are very long and hence this effect does not hinder the performance of
useful experiments. \footnote{A much more exhaustive analysis of this which
largely inspired this work was done by R. Cowsik.}




%
%

\section{The ``polarized" - conducting direction dependent Casimir Effect
at finite temperatures.}

\noindent \subsection{A Path Integral Approach.}

In the standard case of isotropic conducting plates of area $A$, the
Casimir force per unit area is given by \cite{Milton2001}
\begin{equation}
\label{eq:Ken1}
\frac{F_{\sf Casimir}}{A}=-\frac{\pi^2}{240}\frac{\hbar c}{a^4}.
\end{equation}
We now consider the case of plates which conduct only along one direction.

 The exact evaluation of the Casimir energy/force per unit area utilizes
the path integral.  For clarity, we reproduce some of key steps from an
earlier work \cite{Kenneth1998}. Unless explicitly inserted otherwise, we will set  both the speed of light $c $ 
and the Boltzmann constant $k_{B}$ to unity ($c=k_{B} =1$) in the calculations
that follow. In the presence of conducting 
plates 1 and 2, which are both parallel to the x$-$y plane and separated by
a distance ``$a$'' along the $z$ axis the partition function is given by

\begin{eqnarray}
Z = \int D{\cal{A}}~ DJ ~ \exp(-i \int d^{4}x ~ \frac{F_{\mu \nu} F^{\mu \nu}}{4} + i \int
d^{3}x~ {\cal{A}}_{\mu} J^{\mu}),
\end{eqnarray}
\noindent where $F_{\mu\nu}$ are the standard EM fields and $J$ denotes the
currents. In the last term, the scalar product ${\cal{A}}_{\mu}  J^{\mu}$ is
performed only along the area-time of the plates. 
Here and henceforth, $J_{1}$ and $J_{2}$ will denote the currents in the first and second conducting plates, respectively. If both plates conduct
only along a single spatial direction, then $J_{1}$ and $J_{2}$  
will have spatial projection restricted to the $i$th plate (along the specified
conductance directions in each of the plates). The total current $J= J_1+J_2$. 
As is well appreciated, the currents serve as Lagrange multipliers forcing the vanishing of the respective components of
the electric fields along the two plates. Integrating (after a Wick
rotation) the quadratic form in ${\cal{A}}_{\mu}$  we find

\begin{equation}
\label{eq:Ken6}
\int DJ ~\exp \Big[ - \int d^3x~d^3y~\left ( \frac{J_1(x)\cdot
J_1(y)+J_2(x)\cdot J_2(y)}{(x-y)^2} +\frac{2J_1(x)\cdot
J_2(y)}{(x-y)^2+a^2}\right) \Big],
\end{equation}

\noindent where we used the conservation of  $J_1$ and $J_2$  and the
resulting gauge freedom to choose the simple configuration space Feynman
propagator in 4-dimensional space-time, $\Delta^{\mu \nu}_F(x-y) = g^{\mu
\nu}/{2\pi^2(|x-y|)^2}$. It is important to underscore that the currents
$J_{1,2}$, the coordinates, and the momenta $k$ to appear shortly, all live
in the 3-dimensional $(x,y,t)$ space. The first two terms in Eq.
(\ref{eq:Ken6}) refer to the ``self interactions" of currents in the
individual plates and the third to the mutual plate$-$plate interactions -
hence the extra factor of $a^2$ in the denominator of the propagator, with
a the plate spacings in the $z$ direction. Fourier transforming Eq. (\ref{eq:Ken6}) 
which amounts to the (unitary) change of variable from
$J(x)$ to $J(k)$ - we find, thanks to the translation invariance of the
propagator,

\begin{eqnarray}
\label{eq:Ken7}
\int DJ(\vec{k})\exp \Big[ -\int \frac{d^3k}{k}
( J_1(\vec{k})\cdot
J_1(-\vec{k})+J_2(\vec{k})\cdot J_2(-\vec{k}) \nonumber
\\ +2J_1(\vec{k})\cdot
J_2(-\vec{k}) e^{-ka})\Big].
\end{eqnarray}

The conservation $\partial^{\mu}J_{\mu}(x)=0$, or $k \cdot J(k)=0$, of the
currents along with the given angle $\beta$  between their spatial ($xy$)  projections
(along the conductance directions in the two plates) fix the cosine of the
angle between the (three dimensional) $J_1(k)$ and $J_2(k)$ which we denote
by $\alpha(\hat{k})$ and $J_1 \cdot J_2 = J_1J_2 \alpha(\hat{k})$.  The
integration of the quadratic (in $J_i(k)$) action produced the usual
product over modes $k$ of the 2$\times$2 determinants,
\begin{equation}
\label{eq:Ken10}
Z=\prod_{\vec{k}}  \textup{det}\begin{pmatrix}
 \frac{1}{k} &
 \frac{\left (\alpha \hat{k}  \right )}{k} e^{-ka}\\
 \frac{\left (\alpha \hat{k}  \right )}{k} e^{-ka} & \frac{1}{k}
\end{pmatrix}^{-\frac{1}{2}}.
\end{equation}

At zero temperature, taking the log of $Z$, transforming to the continuum limit in the large
volume = $AT$ with $A$ the area of the plates and $T$ the long time
duration, so as to replace the mode sum by an integral, and discarding an
infinite constant that is independent of $a$ (plate separation) and $\beta$ (the angle between the
conductance directions), we find that

\begin{eqnarray}
\label{eq:Ken11}
\ln Z &=&-\frac{1}{2}   \sum_{\vec{k}}\textup{ ln det}\left ( ... \right )
\nonumber \\
                        &=&-\frac{1}{2}AT\int \frac{d^3k}{\left (2\pi
 \right )^3 } \textup{ ln}(1-\alpha^2(\hat{k}) e^{-2ka})+\textup{const}.
\end{eqnarray}

Using  $E=(\ln Z)/ T$, we finally obtain the Casimir energy per unit area

\begin{equation}
\label{eq:Hand01left}
\frac{E}{A}=\frac{1}{2}\int \frac{d^3 k}{\left (2\pi  \right )^3 } \textup{
ln}\left [ 1-\alpha ^2(\hat{k}) e^{-2ka}\right ].
\end{equation}

The three vector $\vec{k}= k \hat{k}$ is, in polar coordinates,

\begin{eqnarray}
\label{eq:Hand01right}
 \vec{k} &=& (k_x,k_y,k_z)= (k_x,k_y, \omega ) \nonumber\\
    &=& k(\textup{sin}\theta \textup{cos}\varphi , \textup{ sin}\theta
\textup{sin}\varphi , \textup{ cos}\theta).
\end{eqnarray}

This yields an explicit expression for $\alpha^2$  as in \cite{Kenneth1998},

\begin{eqnarray}
\label{eq:Hand02}
\alpha ^2&=&\left ( \hat{J}_1(\vec{k}) \cdot  \hat{J}_2(\vec{k})\right )^2
\nonumber \\
               &=& \frac{\left [\textup{cos}\beta -\textup{sin}^2\theta ~
\textup{cos} \varphi~ \textup{cos}(\varphi -\beta )  \right ]^2}{\left
(1-\textup{sin}^2\theta \textup{cos}^2\varphi  \right )\left (
1-\textup{sin}^2 \theta \textup{\textup{cos}}^2(\varphi-\beta) \right )}.
\end{eqnarray}

The last two equations constitute the starting point for the present discussion of
the Casimir force dependence on the angle $\beta$ between the conductance
directions for non-zero temperature T (not to be confused with the
Euclidean time above). The latter manifests by replacing the energy,
$\omega=k_t$ integration above by a sum over the Matsubara frequencies
$\omega(n)= 2\pi n k_{B} T/\hbar \equiv n \tau$

\begin{equation}
\label{eq:Hand03}
\int \frac{dk_t}{2 \pi}\rightarrow T\sum_{-\infty}^{\infty}\left [
\textup{Matsubara frequencies} \right ].
\end{equation}
Using $k =(k_x^2+k_y^2 +\omega^2)^{1/2} \equiv (K^{2} + \omega^{2})^{1/2}$,
 we then have the Casimir free energy per unit area at non-zero temperatures,
\begin{equation}
\label{eq:Hand04}
\frac{{\cal{F}}}{A}=\frac{T}{2}\int  \frac{d^2 K}{\left (2\pi  \right )^2 }
\sum_{n=-\infty}^{\infty}\textup{ ln}\left [ 1-\alpha ^2(\hat{K})
e^{-2a \sqrt{K^2+\left (n \tau  \right )^2}}\right ].
\end{equation}
This last equality and further analysis bellow imply that at high
temperatures (or high Matsubara frequency $\tau$), all $n\neq 0$ terms vanish exponentially. Since the only other
dimensionful energy parameter in the problem is $1/a$ with a the plate
separation, the ``high temperature" dimensionless parameter $\xi$ where
$\xi=2\pi Ta = \omega_1 a$ which is the ratio of the two energy scales must
be large. The key observation is that if we keep only the lowest n=0 mode
we lose $\textit{all}$ dependence on the angle $\beta$ between the
conductance directions!

As seen from Eq. (\ref{eq:Hand01right}), a Matsubara index $n=0$, or
equivalently, $\omega=0$, implies that $\sin\theta=1, \cos\theta=0$.
Substituting Eq. (\ref{eq:Hand02}) we find that $\alpha^2=1$
identically and
\begin{equation}
\label{eq:Hand05}
\frac{{\cal{F}}}{A}|_{n=0}=\frac{T}{2}\int \frac{K~dK} {(2 \pi)^2} \int d\varphi
\textup{ ln}(1- e^{-2Ka}).
\end{equation}

Expanding the integrand in powers of $\exp(-2Ka)$, integrating each
term, and summing up the resulting series $\sum_{l=1}^{\infty}
(1/{l^3})=\zeta(3) \approx 1.202$, we obtain the high temperature limit of
the Casimir free energy: ${\cal{F}}/A \sim T/a^2$. The corresponding Casimir force
is set by $-\frac{d}{da} ({\cal{F}}(a,T))=const~ (T/a^3 )$; this result will be further motivated by
heuristic arguments to be briefly presented later. At high temperatures
($\xi \gg 1$) this ``finite temperature Casimir force'' is larger than that
of the zero-temperature pure case of plates in the vacuum by a factor of
order $\xi$. Our main concern here though is the resulting dramatic
suppression of the dependence on the relative angle $\beta$ between the
conductance direction in the two plates. 
The maximal variation of the Casimir energy as a function of $\beta$ is
obtained as $\beta$ varies from $\beta$=0 to $\beta =\pi/2$, namely
that between parallel and perpendicular conductivities.

In the first, parallel, case we have from Eq. (\ref{eq:Hand02}) that
$\alpha=1$ identically yielding

\begin{equation}
\label{eq:Hand06}
\frac{\cal{F}}{A}|_{ (\beta=0,\textup{ }T)}=\frac{T}{2}\int
\frac{d^2K}{(2\pi)^2}\sum_{n=-\infty}^{\infty}\ln(1-e^{-2a \sqrt{K^2+(n \tau )^2}}),
\end{equation}
at a general temperature $T$.

The case of $\beta=\pi/2$  
is more involved. We have
$\sin^2\theta=K^2/(K^2+\omega^2)=K^2/(K^2+(n\tau)^2)$ and for $n \neq 0$

\begin{equation}
\label{eq:Hand09}
\alpha^2|_{\beta=\frac{\pi}{2}}=\frac{(K^2\textup{ cos} \varphi \textup{
sin} \varphi)^2}{[(n \tau)^2+K^2\textup{ sin}^2\varphi][(n
\tau)^2+K^2\textup{ cos}^2\varphi]}<1.
\end{equation}

This should be substituted in

\begin{equation}
\label{eq:Hand10}
\frac{{\cal{F}}}{A}(\beta=\frac{\pi}{2},T)=\frac{T}{8\pi^2}\int d\varphi\int KdK\sum_{n=-\infty}^{\infty} \textup{ln}  \left [1-\alpha^2_{(\beta=\frac{\pi}{2})}e^{-2a \sqrt{K^2+(n\tau)^2}}  \right ].
\end{equation}

It is difficult to separate the plates in the parallel planar geometry
required here, by less than a micron, so that $a\geq10^{-4}$ cm. At room
temperature, $T=300$ K, we find already for this small separation $a$ that
largest terms with non-trivial angular dependence, i.e., the $n = \pm 1$ terms, are small.
For $a=1$ micron, $2 \alpha^{2}_{\beta = \pi/2} \exp(-2 \sqrt{K^{2} + (2 \pi a T)^{2}}) < 2 \exp(-4 \pi k_{B} T a/(\hbar c)) \approx 0.353$. The largest contribution to the Casimir free energy and force originates from the first ($n=0$) term in the Matsubara sum which is trivially angle ($\beta$) independent ($\alpha^{2}(n=0) =1$). The $\beta$ dependence originating from the non-zero Matsubara frequencies is reduced by a factor of three for a plate separation of a micron (and is exponentially decreasing in $a$). Reserving a more careful evaluation of
this effect at various temperatures to future work, we still wish to have a
better physical understanding as to why the rather strong $\beta$
dependence at $T$=0 rapidly diminishes as the temperature is increased. To this end, we
briefly review heuristic albeit crude arguments that seems to suggest that
the $\beta$ dependence is not altogether removed.

\subsection{The heuristic approach, thermal decoherence,  and its
limitations.}

\begin{figure} [ht]
\center
\includegraphics[width=12cm]{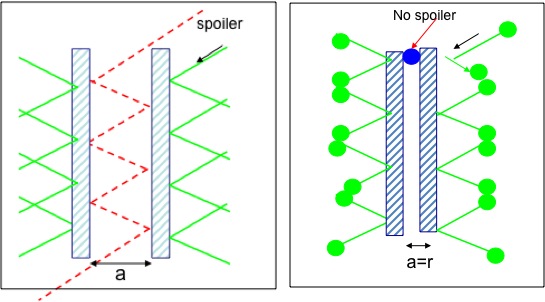}
\caption{A cartoon of a classical analogue of the Casimir effect.
Left panel: A cancellation of outside and internal forces leading to no net
attraction between the plates.
               Right panel: If the gas particles are of a "quantized"
minimal size, there are no internal forces and thus an overall attraction
between the plates when they are close to each other.}
\label{fig:FigSlide3}
\end{figure}

Forces between macroscopic objects are often the consequence of interaction
with smaller particles and/or excitations in the surrounding medium. For
our purpose it is useful to consider then two parallel plates immersed in a thermal
molecular gas that are a distance $a$ apart. One might think that the molecules impinging on the plates
from the outside there will generate a net pressure pushing them towards
each other. This conflicts with basic expectation that in a system with
uniform pressure \textit{no} net force can be exerted on any object. The
nice resolution is shown in the left panel of Figure (\ref{fig:FigSlide3})  showing that the
rare ``spoiler'' molecules that sneak in between the plates keep moving back
and forth and reflecting from the plates exactly counter the effect of all
the many outside collisions. If, however, as shown in the right panel of Figure
(\ref{fig:FigSlide3}), the gas molecules have a distribution of sizes and
in particular have diameters of order a then they will not be able to go
inside and a net apparent attraction between the plates will thus be
generated.

It seems natural to extend these simplistic arguments - which indeed have
been verified in colloidal systems - to a gas of thermal photons. Indeed
photons/$e.m$ modes of wavelengths $\lambda$ exceeding the plate separation
$a$ - cannot propagate inside this ``wave guide''. Such modes cannot serve
then as ``spoilers" for the inward pressure from the outside collisions.
The number density of such modes of wavelength larger than  $\lambda
\approx {a}$ is of the order of $a^{-3}$. In the 
$k_{B} T> \hbar c/a$ regime, each mode
has an energy of $k_{B}T$ (and each mode is 
is populated by $k_{B}T/(\hbar \omega) \approx k_{B} T a/(\hbar c) \approx \xi$ quanta).
The resulting net pressure
and attractive force paper unit area $F_{\sf Casimir}(a,T)/A$ is $\propto
T/a^3$ as predicted by the path integral formulation above.

It is well known that many phenomena in Q.E.D derive and/or can be
attributed to the existence of the vacuum fluctuations with each mode
carrying ``half'' a quantum or zero point energy of $\hbar  \omega/2$. A
simple repetition of the above leads to $F_{\sf Casimir}(Vac) \approx 1/{a^4}$.
However when pushed further to more subtle effects such as the dependence
of the force on the angle ${\beta}$ between the directions of conductivity
of the two plates which is the topic of interest here, the naive argument
seems to completely fail.

\begin{figure} [ht]
\center
\includegraphics[width=12cm]{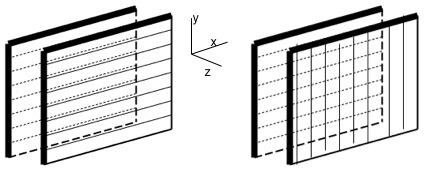}
\caption{Plates that conduct along one direction for different angular
orientations of the conducting directions. Left panel: Parallel plates
($\beta =0$).
               Right panel: Orthogonal plates ($\beta = \pi/2$).}
\label{fig:FigSlide4}
\end{figure}

To see this let us consider the following setup which hopefully can be used
to heuristically motivate and in some subtle incarnation also measure the
effect. Instead of having the parallel Casimir plates made of full metal
sheets we have two frames strung with sufficiently dense and thin parallel
wires. These then correspond to the Casimir plates with the preferred
conductance directions namely along the wires in each frame. The two
extreme cases studied above then correspond to the setting of Figure
(\ref{fig:FigSlide4}).

In the first case only photons polarized along the common - say $x$
direction - of the wire will be effectively reflected by the induced
currents in the thin wires. We then expect that the Casimir force in such
an arrangement will be reduced by a factor half as and amusingly this was
indeed found both for the $T=0$ and for the $T \neq 0$ cases by the path
integral method. If, however, the wires are crossed at 90 degrees then the
polarized photons coming from the outside will sail through the left mesh
and then reflect back from the right mesh . However since an equal amount
of y polarized photons reflect from the outside of the right mesh - no net
force should arise. This is flatly negated by the path integral
computations.

First it was found \cite{Kenneth1998} that for the $T=$0 case, the force in
the crossed case, as illustrated on the right panel of Figure (\ref{fig:FigSlide4}), is reduced by only by
$\approx 1/2$ as compared with the parallel wires case, depicted in the lefthand panel of Figure
(\ref{fig:FigSlide4}a). This can be `explained' by the fact that our
argument naively assumed that all photons move along the $z$ direction
whereas all direction of incidence should be considered. The key question
still remains - why did we go so wrong in the thermal case and what makes
here the direction/polarization effect 
decay so fast- as $\exp(-\xi) = \exp(-2 \pi k_{B} Ta/(\hbar c))$?

We believe that the answer is relatively simple though instructive. The
naive argument fails because of the decoherence effects at finite
temperature. Clearly the path integral calculation which is far more
sophisticated seems superior and is indeed correct. Specifically we expect
that practically at all times the photons are not in any specific
polarization state but in an equal mixture of both.

The exponential suppression of the effect with temperature could be
qualitatively argued for as follows: the average number of photons in any
mode $k$ of frequency $\omega$ is $\bar{n}= (k_{B}T)/ (\hbar \omega)
\approx \xi$ The probability of having zero thermal photons then is
$\exp(-\bar{n})$. Since all thermal photons are maximally mixed we have
then only the ``1/2'' vacuum photon contributing the exponentially suppressed
effect.

 Let us assume that the decoherence time is set just by the temperature and
is indeed $t_{dec} = h/(k_{B}T)$. If the distance travelled during this
time - the decoherence distance $l_{dec} = (ct_{dec})$ is much larger than
the separation of the plates $a$ than while in transit the photon could
decohere $\xi= a/\ell_{dec}$ times. The probability that it will stay in its
original polarization state and contribute to the angular ($\beta $)
dependent force is then reduced by $\exp (-\xi)$.

 The key observation is that, up to constants of order unity,  $t_{dec} =
h/(k_{B}T)$ is the shortest minimal decoherence time which is consistent
with the quantum uncertainty relation:
$\Delta t \Delta E \ge \hbar/2 $ when 
we substitute the thermal energy ($k_{B} T)$ for the energy uncertainty scale $\Delta E$. 
In specific set-ups,we might have it much longer
than that this lower bound scale.
 
It is worth noting that the formal path integral approach in which the
attraction between say two charged particles can be viewed as resulting
from the joint propagation in the same background field is not completely
different from what we are discussing albeit in a very simplistic view
here. What the naive, almost mechanistic, approach sorely misses are
Aharonov-Bohm (or even Coulomb) phases of the form
$\exp{\int{dx^{\mu} {\cal{A}}_{\mu}(x)}}$.

We next turn to the effects of thermal decoherence
on Casimir-like forces generated by such Aharonov-Bohm (AB)
phases.

\section{Thermal decoherence screens AB phase induced interactions}
\label{sec:4}

The better known evaluations of the Casimir energy are by the summing  the
energy shifts of all the photon modes in the vacuum, induced by the
introduction of the plates at their specified location. It is assumed that
the plates or other conducting/ dielectric objects, have large inertia and
can be viewed as static. Some time ago it was pointed out \cite{Aharonov0000a}
that similar ``AB type" interactions are generated between  magnetic fluxes
$\Phi_1,\Phi_2,...,\Phi_{n_F}$  located at given points $ R_1,...,R_{n_F}$
inside a two dimensional region where the wave function of N electrons is
non-vanishing. The introduction of the fluxes which play here the role of
the heavy degrees of freedom modifies the the ground state wave-function:

 \begin{eqnarray}
 \Psi^{(0)}(\vec{r}_{1}, \vec{r}_{2}, \cdots, \vec{r}_{N}) \to
\Psi^{(0)}(\vec{r}_{1}, \vec{r}_{2}, \cdots, \vec{r}_{N}; \vec{R}_{1},
\Psi_{1}, \vec{R}_{2}, \Psi_{2}, \cdots, \vec{R}_{n_{F}}, \Phi_{n_{F}}).
 \end{eqnarray}

The energy shift relative to the magnetic field free case, $\delta
E^{(0)}$, generated by having magnetic fluxes depends on their sizes and
location. This shift generates via a Born Oppenheimer type approximation,
an interaction between the fluxes

\begin{eqnarray}
\delta E^{(0)} (\vec{R}_{j}, \Phi_{j}) = W(\vec{R}_{j}, \Phi_{j}),
\end{eqnarray}

\noindent where the gradients of $W$ are the forces $\vec{F}_{i}$ that act
on the fluxes $\Phi_{i}$. In the following we recall the estimate of this
interaction and resulting forces between the fluxons so as to be able to
discuss the thermal decoherence induced screening of theses forces. The
calculations are greatly simplified if we first neglect the interaction
between the electrons so that the ground state is
antisymmetrized product state (filling up a Fermi circle in an appropriate
parametrization). The total energy then is simply the sum of the shifts of
the individual states

 \begin{eqnarray}
 \delta E^{(0)}(\vec{R}_{j}, \Phi_{j}) = \sum_{i=1}^{N} \delta
E_{\gamma_{i}} (\vec{R}_{j}, \Phi_{j}).
 \end{eqnarray}

A sense of the size of the effect is obtained by considering first just one
flux $\Phi=(\alpha \Phi_0)$  at the center of a circular disc of radius R
with a uniform two dimensional electron density $n_2$. The fluxon $\Phi_0 =
hc/e$ is the flux quantum appropriate to the charge e of a single electron.
By rotational symmetry, the initial  electron wave functions are
eigenstates of the angular momentum $L_z$ with integer eigenvalues $l=0,
\pm 1, \pm 2,...$. For the purpose of finding the effect of the fluxes, the
exact $J_n^l(kr) \exp{il\phi}$ form for the wave-functions, can be WKB
approximated by semiclassical paths of fixed radii. The introduction of the
fluxon increases all of the $l$ values by  $\alpha$,

\begin{eqnarray}
|l| \to |l| + \alpha \mbox{~for~} l \ge 0, ~ ~ |l| \to |l| - \alpha
\mbox{~for~} l <0,
\end{eqnarray}

\noindent and correspondingly modifies the relevant angular parts of the
energies  $ E_{l,r}^0=|l|^2 \hbar ^2/(2mr^2)$.

The sum of the energy shifts for a single  $ l=+,- |l|$ pair is then
shifted by $\delta E^{(0)}_{|l|,r} = \frac{\hbar^{2} \alpha^{2}}{2
mr^{2}}$. Summing over all l and r values so as to account for all the
$N=n_2\pi R^2$ electron states we find that the total energy shift is:
\begin{eqnarray}
W_{\alpha}^{tot} = \sum_{|l|,n} \frac{\hbar^{2} \alpha^{2}}{2 mr_{n}^{2}},
~ ~  \frac{\alpha^{2}}{2}
\frac{n_{2} \hbar^{2}}{2m} \int_{0}^{R} \frac{2 \pi r ~dr}{r^{2}} =
\frac{\pi}{2} \frac{\alpha^{2} n_{2} \hbar^{2}}{2m} \ln (\frac{R}{a_{0}}),
\end{eqnarray}
\noindent where $a_0$, the distance betweeen the electrons  serves as a
lower cutoff.

 The logarithmic dependence of $W(R)$ on the size of the system reflects an
underlying scaling invariance of the effective two-dimensional potential
generated by the fluxes. It suggests that the interaction between two
fluxons $\Phi_1$ and $\Phi_2$ at a distance $a=|R_1-R_2| \ll{R}$ and far
from the boundaries is:

 \begin{eqnarray}
 W(\alpha_{1}, \alpha_{2}) = \xi_{(\alpha_{1}, \alpha_{2})}  \frac{\pi}{16}
\frac{n_{2} \hbar^{2}}{m} \ln (\frac{a}{a_{0}}),
 \end{eqnarray}

\noindent  with  $\xi_{(\alpha_1,\alpha_2)}$ some function of the
individual fluxes and a logarithmic dependence on the relative distance.

Unlike the Casimir force,  the present force is of a toplogical nature. The
energy $W_{\alpha}(R)$ due to a fluxon $\alpha$ is proportional to
$\alpha^2$ only for $\alpha \le{1/2}$. Integer $\alpha$ amount to pure
gauge as a  shift of all angular momenta by an integer amounts to a
negligible "surface" effect modifying the energy of states with angular
momenta near  $l_{max}$ only. This and the time reflection symmetry
corresponding to $L_z \rightarrow{-L_z}$ imply that $W_{\alpha}(R)$ is
maximal for $\alpha=1/2$ and falls off as  $(1-\alpha)^2$ in the $[1/2,1]
$interval and periodically repeats beyond that. For this reason the
pair-wise force is maximal  between fluxes of size $|\alpha| = 1/2$.
Furthermore, the force is attractive even when the latter are parallel to
each other since as the fluxons move closer together more electrons see
just the effective total and trivial integer flux and pick no phase upon
encircling the pair:
\begin{eqnarray}
 F_{(1/2,1/2)}(a) \simeq \frac{\xi \pi}{16} \frac{n_{2} \hbar^{2}}{m}
 \frac{1}{a}.
 \end{eqnarray}

 The topological nature also manifests in that no force is experienced by a
fluxon located outside the region where the electrons are confined since
the phases picked up by the electrons when traveling along  any closed loop
-which drive all these effects,- do {\textit{not}} depend on the location
of out-side fluxon.

Half fluxons that are parallel to each other are particularly relevant as
they may naturally occur in superconductors. Unfortunately this makes
calculations for assessing the observability of  this new type of force
much harder than for the case of the formally somewhat similar Casimir
effect.
 The Casimir plates are well defined macroscopic objects whereas the
fluxons are generated via a collective effect of the Cooper pairs and
electrons in the superconductor, i.e., the basic element comprising the
medium which generates the new force between the fluxons.
It was conjectured \cite{Aharonov0000a} that the large mutual interaction
energies of the fluxons might induce some  counter-currents of electrons
which weaken the other-wise very strong force.

 Here we  focus on the more concrete thermal decohering which can reduce
the force. It too is connected with the specific material effects. In the
case of the Casimir plates one could reduce the ambient temperature to very
low values so that (when all constants are restored) the dimensionless product $\xi = a k_{B} T/(\hbar c)$ is small enough and
 also the path integral formal calculation will allow a measurable angle
dependent Casimir forces. This however cannot be done here even in
principle. The temperature $T$ of the superconductor where we propose to
study possible attraction between fluxons {\it{must}} be a finite fraction
$f = {\cal{O}}(1)$ of the critical temperature for the onset of
superconductivity, $ T=f T_c$. Only then will a finite fraction $u= F(f)$
of all the electrons remain unpaired into cooper pairs, and only these
unpaired electrons generate the force between the half-fluxons.

 Thus we have to  address the question of the thermal decoherence effects
on the new force between the fluxons. In order to disentangle this from
other matter linked effects we  use the idealization where the fluxons can
be viewed as external rigid entities immersed into the electrons. We could
then compute the ``Casimir like force" per unit length  due to the
electrons in the medium by using a path integral and then include the
thermal effects by compactifying the euclidean time direction to the
Distance 1/T.
Two technical difficulties may  hinder such a calculation.  First, even for
non-interacting electrons, the path integral has the sign problem arising
from the Fermi statistics and the non-vanishing chemical potential. Also
the symmetry of the problem is reduced from the 3D translational invariance
(in the $x,y$ and $t$ directions) when we have the parallel infinite Casimir plates
geometry to one with only a spatial invariance with $z$ the direction of the
fluxons. \cite{foot1}

We will not attempt such a calculation here.
We believe, however, that at $T=0$, it will reveal the logarithmic interaction
energy of  \cite{Aharonov0000a}. Also as in the path integral calculation of the
Casimir force, at finite T, this long range interaction will be screened
and fall exponentially with a characteristic decoherence length
$l_{dec}=l_{scr}$ . It derives from the decoherence time $t_{dec}=1/T $
 for $T \approx{T_c}$ via $l_{dec}=t_{dec} v_{Fermi}$ as the latter Fermi
velocity is the velocity of the electrons.
For the force to be readily measurable we need that the
average distance L between the parallel fluxes in an Abrikosov 
lattice
be smaller than $l_{scr}$. But L  must be greater than the radius d of the
individual fluxons which, in turn, is of the order of the London
penetration length  $\lambda_L$. The condition then becomes $l_{coh}
>\lambda_L$ whereas the opposite inequality characterizes the type II
superconductors in which Abrikosov flux lattices exist . Indeed  the very
existence of such lattices requires that the ordinary magnetic repulsive
force between the parallel fluxons will dominate.
\cite{foot2}

 \section{Summary and conclusions}

In the current work, we discussed  two quantum systems (involving the Casimir and  AB type effects) 
 examined their stability to thermal decoherence. 
 
 In the first case, an explicit path integral calculation for the
finite temperature system, suggested substantial decoherence for the
polarized Casimir force rendering experimental attempts for
measuring the relative conductance angle dependence force unlikely to succeed.

We have not been able to perform an explicit $T=0$ and $T\ne{0}$ path
integral evaluation of the second system with AB force between fluxons in a
superconductor
reflecting the phases (sign flip) picked up by unpaired electrons
circulating the fluxons in superconductors. However similar arguments to
those used
in the Casimir case suggest that the situation may be notably
different. 

The conventional coherence length for electrons $\hbar v/(k_{B} T_{c})$ is smaller in
generic type II superconductor than the London
penetration length which in turn is smaller than the diameter of a single
fluxon let alone the distance between two fluxons. At temperatures $T\approx T_c/2$,
the formal minimal de-coherence length which translates into an exponential
screening of the original long range force at $l_{scr}=l_{dec}$ will then
dramatically reduce the new force and make its detection rather difficult.

 There are many other obstacles of a more technical nature for observing
this truly unique and interesting force such as the existence of
impurities
 and the possible pinning of the fluxons on those. Still it is not
inconcievable that a dedicated high precision measurements of Abrikosov
lattices for disparate superconductors and temperatures in which
both the fraction of unpaired electrons and the (de)coherence
length
and of the magnetic field B (changing the density of fluxons) may reveal the
existence of the force.
 While differing greatly in many details both of the above examples
illustrate albeit the very different, effects of decoherence introduced by
finite
 temperature and both serve as concrete actual examples of the much more
formal yet general discussions of the effects of strongly and weakly
coupled thermal baths on quantum systems.

\begin{acknowledgements}
We would like to thank Armin Gulian and Michael Levin for insightful and very
helpful comments and R. Cowsik for many discussions over the years and in
particular a recent one on classical dissipation which inspired the present
work. This work was partially supported by NSF DMR-1106293.
\end{acknowledgements}


\begin{thebibliography}{}
\bibitem{foot1} To see thermal screening it suffices to consider a single
fluxon in which case we can use, as done above,  the extra $\phi$ rotational
symmetry to help facilitate the calculation.
\bibitem{foot2} Note that  in the present case we have no good reason to
suspect that the coherence time will be longer
than the minimal scale $h/(k_{B} T)$: the medium of electrons is dense and we have a very strong
electron electron scattering cross-section. Even upon traveling a
distance of, e.g., 300 Angstrom between fluxons in a NbSe$_2$ superconductor, an
electron is likely to collide with other electrons and randomize its
phase.
\end{thebibliography}


\end{document}